\documentclass[aps,pra,showkeys,twocolumn,superscriptaddress]{revtex4}
\usepackage{array}
\usepackage{booktabs}
\usepackage{tabu}
\usepackage{dcolumn}
\usepackage{amsmath}
\usepackage{amsfonts}
\usepackage{float}
\usepackage{amssymb}
\usepackage{graphicx,color}
\usepackage[colorlinks={true}]{hyperref}
\hypersetup{colorlinks=true,linkcolor=red,citecolor=blue,urlcolor=blue}
\usepackage{graphicx}
\usepackage{subfigure}
\usepackage{graphicx}% Include figure files
\usepackage{dcolumn}% Align table columns on decimal point
\usepackage{bm}% bold math
\usepackage{pstricks}

\def\be{\begin{equation}}
  \def\ee{\end{equation}}
\def\bea{\begin{eqnarray}}
\def\eea{\end{eqnarray}}
\def\f{\frac}
\def\n{\nonumber}
\def\l{\label}
\def\p{\phi}
\def\o{\over}
\def\R{\rho}
\def\pa{\partial}
\def\om{\omega}
\def\na{\nabla}
\def\P{\Phi}
\begin{document}
\title{Does conditional entropy squeezing indicate normalized entropic uncertainty relation steering?}
\author{A-S. F. Obada}
\affiliation{Mathematics Department, Faculty of Science, Al-Azher University, Nassr City 11884, Cairo, Egypt}
\author{M. Y. Abd-Rabbou}\email{m.elmalky@azhar.edu.eg}
\affiliation{Mathematics Department, Faculty of Science, Al-Azher University, Nassr City 11884, Cairo, Egypt}
\author{Saeed Haddadi}\email{saeed@ssqig.com}
\affiliation{Faculty of Physics, Semnan University, P.O.Box 35195-363, Semnan, Iran}
\affiliation{Saeed's Quantum Information Group, P.O.Box 19395-0560, Tehran, Iran}
\date{\today}% It is always \today, today,
\def\be{\begin{equation}}
  \def\ee{\end{equation}}
\def\bea{\begin{eqnarray}}
\def\eea{\end{eqnarray}}
\def\f{\frac}
\def\n{\nonumber}
\def\l{\label}
\def\p{\phi}
\def\o{\over}
\def\R{\rho}
\def\pa{\partial}
\def\om{\omega}
\def\na{\nabla}
\def\P{$\Phi$}
%\nofiles

%=============================================================%
%=============================================================%
%============== Abstract =======================================%
%=============================================================%
%=============================================================%
\begin{abstract}
A novel approach is introduced to assess one-way Normalized Entropic Uncertainty Relation (NEUR)-steering in a two-qubit system by utilizing an average of conditional entropy squeezing. The mathematical expressions of conditional entropy squeezing and NEUR-steering are derived and presented. To gain a better understanding of the relationship between the two measures, a comparative analysis is conducted on a set of two-qubit states. Our results reveal that the two measures exhibit complete similarity when applied to a maximally entangled state, while they display comparable behavior with minor deviations for partially entangled states. Additionally, it is observed that the two measures are proportionally affected by some quantum processes such as acceleration,  noisy channels, and swapping. As a result, the average of conditional entropy squeezing proves to be an effective indicator of NEUR-steering.
\end{abstract}


\maketitle

%%%%%%%%%%%%%%%%%%%%%%%%%%%%%%%%%%%%%%%%%%%%%%%%%%%%%%%%%%%%%%%%%%%%%%%%%%%%
%%%%%%%%%%%%%%%%%%%%%%%%%%%%%%%%%%%%%%%%%%%%%%%%%%%%%%%%%%%%%%%%%%%%%%%%%%%%
%%%%%%%%%%%%%%%%%%%%%%%%%%%%%%%%%%%%%%%%%%%%%%%%%%%%%%%%%%%%%%%%%%%%%%%%%%%%
%%%%%%%%%%%%%%%%%%%%%%%%%%%%%%%%%%%%%%%%%%%%%%%%%%%%%%%%%%%%%%%%%%%%%%%%%%%%
%============  Sec.I (Introduction)  =======================================
%%%%%%%%%%%%%%%%%%%%%%%%%%%%%%%%%%%%%%%%%%%%%%%%%%%%%%%%%%%%%%%%%%%%%%%%%%%%
%%%%%%%%%%%%%%%%%%%%%%%%%%%%%%%%%%%%%%%%%%%%%%%%%%%%%%%%%%%%%%%%%%%%%%%%%%%%
%%%%%%%%%%%%%%%%%%%%%%%%%%%%%%%%%%%%%%%%%%%%%%%%%%%%%%%%%%%%%%%%%%%%%%%%%%%%
%%%%%%%%%%%%%%%%%%%%%%%%%%%%%%%%%%%%%%%%%%%%%%%%%%%%%%%%%%%%%%%%%%%%%%%%%%%%
\section{INTRODUCTION}	%) A SECTION HEADING
In 1935, Schr\"{o}dinger tried to interpret the Einstein-Podolsky-Rosen (EPR) paradox by establishing correlations between two quantum systems that were too strong to be explained classically, this phenomenon is commonly referred to as EPR-steering \cite{PhysRev.47.777,schrodinger1935discussion}. The concept of steering involves one remote user using a pair of entangled states to influence or steer their partner's state through local measurements. As per the hierarchy of quantum correlations, steerable states are a strict superset of the states that can demonstrate Bell nonlocality and a strict subset of the entangled states \cite{PhysRevLett.98.140402,PhysRevA.92.032107,andp2023}.
Quantum steering has recently received significant attention in the field of quantum information research and has been the subject of both experimental and theoretical investigations \cite{wittmann2012loophole,RevModPhys.92.015001}. For example, the experimental quantum steering has been studied through the implementation of generalized entropic criteria and dimension-bounded steering inequalities, where two or three measurement setups are used on each side \cite{PhysRevLett.125.020404}. Steering game based on the all-versus-nothing criterion has been experimentally demonstrated \cite{PhysRevLett.113.140402}. The asymmetric property of EPR steering is relevant for experimental and potential applications in quantum information as a one-sided device-independent quantum key distribution \cite{PhysRevA.85.010301}, quantum teleportation \cite{PhysRevLett.115.180502}, and optimal prepare-and-measure scenarios \cite{passaro2015optimal}. Moreover, for different quantum systems, the possibility of quantum steering is experimentally interpreted, including photon polarizations in a linear-optical setup \cite{bartkiewicz2016experimental}, Bohmian trajectories \cite{xiao2017experimental}, a family of the natural two-qubit state \cite{PhysRevResearch.4.013151}, and non-Gaussian state \cite{PhysRevLett.128.200401}.

In the theoretical framework, researchers have developed asymmetric criteria of steering correlation for a pair of arbitrary continuous variables \cite{PhysRevLett.98.140402}. Additionally, Walborn et al. \cite{PhysRevA.87.062103} have utilized the entropic uncertainty relations to express the steering inequality for arbitrary discrete observables.  The violation of the Clauser-Horne-Shimony-Holt inequality has also been employed to discuss the degree of steerability \cite{cavalcanti2015analog,ruzbehani2021simulation}. Furthermore, some investigations have been conducted on the violation of steering inequality and its degree for various quantum systems, including a three-mode optomechanical system \cite{PhysRevA.89.022332}, Heisenberg chain models \cite{chen2018quantum,li2018quantum},  two-level or three-level detectors \cite{liu2018influence,wang2018generation}, and qubit-qubit as well as qubit-qutrit states \cite{sun2017exploration,abd2022improving,rahman2023bidirectional}.

On the other hand, the essential conceptions of squeezed spin systems were introduced by Kitagawa and Ueda in 1993 \cite{PhysRevA.47.5138}. The entropy squeezing for a bipartite system has been obtained for three discrete observables in $N$-dimensional Hilbert space and employing the discrete Shannon entropy \cite{fang2000entropy}. The violation of two quadratures of entropy squeezing inequality represents a magnificent indicator of entanglement \cite{el2008single}. Meanwhile, the entropy squeezing of multi-qubit inside a cavity system has been considered a hot research topic, such as: two-qubit interacting with two-mode cavity field \cite{khalil2020entanglement}, qutrit state in a cavity filed \cite{liu2019entropy}, and the effect of classical field and non-linear term on the qubit-field interaction \cite{obada2018influence}.

Our motivation is to introduce how entropy squeezing can be employed as an indicator of the degree of steerability. Overall, as the discrete conditional Shannon entropy is used as a measure of steerability, so do the two quadratures of conditional entropy squeezing express the steering?
This paper is organized as follows. In Section \ref{S1}, we
present the steerability based on conditional entropy squeezing. In Section \ref{S2}, the main results of our paper are discussed in detail. Finally, the conclusion is given in Section \ref{S3}.

%%%%%%%%%%%%%%%%%%%%%%%%%%%
\section{Steerability based on conditional entropy squeezing}\label{S1}
In order to gain a better understanding of the relationship between entropy squeezing and normalized entropic uncertainty relation (NEUR)-steering  for bipartite subsystems $A$ and $B$, we can take advantage of the definition provided by Walborn et al. \cite{PhysRevA.87.062103}. The mathematical framework of NEUR-steering inequality concerning an even $N$-dimensional Hilbert space along with the local hidden state for a pair of arbitrary discrete observables is expressed as \cite{PhysRevA.87.062103}
\begin{equation}
	\sum_{i=1}^{N+1}H(R^B_i|R^A_i)\geq \frac{N}{2} \ln(\frac{N}{2})+(1+\frac{N}{2}) \ln(1+\frac{N}{2}),
\end{equation}
where $ \{R^A_i\}$ and $\{R^B_i\} $ are the eigenvectors of the discrete observables $ \hat{R}^A $ and $ \hat{R}^B $, respectively, and $ N $ is the total number of different eigenvectors. $H(R^B|R^A) \geq \sum_{\lambda} P(\lambda) H_Q(R^B|\lambda) $ denotes the corresponding local hidden state constraint for discrete observables, which is defined by the conditional information entropy $ H_Q(R^B|Q) $ of the probability distribution $ P_Q(R^B|\lambda) $ with the hidden variable $\lambda$.  In two-dimensional Hilbert space $  N =2  $, by employing the Pauli spin operators $\{\sigma_x,\sigma_y,\sigma_z\}$ as measurements, the NEUR-steering from $ A $ to $ B $ is realized only if the following condition is violated \cite{PhysRevA.87.062103,sun2017relativistic}
\begin{equation} \label{e1}
H(\sigma_x^B|\sigma_x^A)+H(\sigma_y^B|\sigma_y^A)+H(\sigma_z^B|\sigma_z^A)\geq 2\ln2,
\end{equation}
where
\begin{equation}
	\begin{split}
		H(\sigma_i^B|\sigma_i^A)&= H(\hat{\rho}_{AB})_i - H(\hat{\rho}_A)_i\\&
		= - \sum_{n,m=1}^{2} P_i^{n,m} \ln P_i^{n,m}+ \sum_{l}^{2} P_i^{l} \ln P_i^{l}.
	\end{split}
\end{equation}
Here, $ P_i^{n,m}=\langle \phi^i_n, \phi^i_m|\rho_{AB}|\phi^i_n, \phi^i_m \rangle $ and $ P_i^{n}= \langle \phi^i_n |\rho_{A}|\phi^i_n \rangle  $ are the probability distribution of an arbitrary two-qubit state $ \rho_{AB} $ and reduced single qubit state $ \rho_{A} $, respectively, where $ |\phi^i_j \rangle  $ represent the two possible eigenvectors ($ j=1,2 $) of $ \sigma_i $, and $ \rho_{A}=Tr_B[\rho_{AB}] $.
\par
 In this paper, we assume that the density state $ \hat{\rho}_{AB} $ with real components in the standard basis $ \{|00\rangle, |01\rangle, |10\rangle, |11\rangle \} $ can be written as
\begin{equation} \label{xxx}
	\hat{\rho}_{AB}=\begin{pmatrix}
\rho_{11}&  0 & 0& \rho_{14}\\
 0&\rho_{22}& \rho_{23}& 0\\
0& \rho_{23}&\rho_{33} & 0\\
\rho_{14}& 0& 0& \rho_{44}
	\end{pmatrix}.
\end{equation}
Note that the operator $\hat{\rho}_{AB}$ satisfies the common conditions  $\hat{\rho}_{AB}\geq 0$ and $Tr[\hat{\rho}_{AB}]=1$.

By applying state (\ref{xxx}) in Eq. (\ref{e1}) and violating NEUR-steering inequality, one can obtain
\begin{equation}
	\begin{split}
		\mathcal{I}_{AB}= & \sum_{i=1}^{3} \sum_{j=1}^{4}\frac{1 + x_{ij}}{2} \ln (1+x_{ij})\\ &- \sum_{k=1}^{2}(1 + a_{k}) \ln (1+a_{k})\leq 2 \ln 2,
	\end{split}
\end{equation}
where the summations in the first term are related to the three Pauli spin operators and probability distribution of the two-qubit $ \rho_{AB} $, respectively, and $ x_{ij} $ are obtained by
\begin{equation}
	\begin{split}
		&x_{11}=x_{12}=-x_{13}=-x_{14}=2 (\rho_{14}+\rho_{23}),\\&
		x_{21}=x_{22}=-x_{23}=-x_{24}=2(\rho_{23}-\rho_{14}),\\&
		x_{31}=3\rho_{11}-(\rho_{22}+\rho_{33}+\rho_{44}), \\ &x_{32}=3\rho_{22}-(\rho_{11}+\rho_{33}+\rho_{44}),\\& x_{33}=3\rho_{33}-(\rho_{11}+\rho_{22}+\rho_{44}),\\
  &x_{34}=3\rho_{44}-(\rho_{11}+\rho_{22}+\rho_{33}).
	\end{split}
\end{equation}

Likewise, the summation in the second term is related to the probability distribution of the reduced state $ \rho_{A} $, and $ a_{k} $ is given by
\begin{equation}
	\begin{split}
		a_{k}=(-1)^k (\rho_{11}+\rho_{22}-\rho_{33}-\rho_{44}).
	\end{split}
\end{equation}

However, the one-way NEUR-steering is quantified based on observable $ A  $ measurements as follows \cite{PhysRevA.93.020103}
\begin{equation} \label{sab}
	S^{A\longrightarrow B}=\max\bigg\{0,\frac{\mathcal{I}_{AB}-2\ln2}{\mathcal{I}_{max}-2\ln2}\bigg\},
\end{equation}
where $\mathcal{I}_{max}= 6 \ln2 $ when the system is prepared in Bell states.

On the other hand, if we define the function $ \Xi(\sigma_i^B|\sigma_i^A)  =  e^{H(\sigma_i^B|\sigma_i^A)} $, then the inequality (\ref{e1}) can be reformulated as
\begin{equation}
	\Xi(\sigma_x^B|\sigma_x^A) \Xi(\sigma_y^B|\sigma_y^A) \geq\frac{4}{\Xi(\sigma_z^B|\sigma_z^A)},
\end{equation}
where
\begin{equation}
	\Xi(\sigma_i^B|\sigma_i^A)= \sum_{n,m=1}^{2} \bigg(P_i^{n,m}\bigg)^{P_i^{n,m}} \times \sum_{l=1}^{2} \bigg(P_i^{l}\bigg)^{P_i^{l}}.
\end{equation}
According to Ref. \cite{fang2000entropy}, the fluctuations in component $ \Xi(\sigma_i^B|\sigma_i^A)  $ ($ i=x,y $) are
said to be ``squeezed in entropy" if the squeezing factor $ E(\sigma_i^B|\sigma_i^A) $ satisfies the condition
\begin{equation}\label{con}
	E(\sigma^B_i|\sigma^A_i)=\max\big\{ 0,\frac{2}{\sqrt{\Xi(\sigma_z^B|\sigma_z^A)}}-e^{\Xi(\sigma_i^B|\sigma_i^A)}\big\},
\end{equation}
with $ i=x,y $. From the previous condition, we can depict the upper bounds or the lower bounds of the NEUR-steering degree.  If the state is a maximum entangled state, then the upper and lower bounds in condition (\ref{con}) are identical. Hence, bidirectional steerability and the average of conditional entropy squeezing quadrature have similar behavior. In partially entangled states, the  NEUR steerability is restricted between the upper and lower bounds. Therefore, the average of the two components of conditional entropy squeezing $ E(\sigma_i^B|\sigma_i^A) $ represents an indicator for quantum steerability. In any case, we can define the quantum steerability based on the average of entropy squeezing as
\begin{equation}\label{ZZ}
	\mathcal{Z}^{A\longrightarrow B}=\max \big\{0,\frac{E(\sigma_x^B|\sigma_x^A)+E(\sigma_y^B|\sigma_y^A)}{2}\big\},
\end{equation}
where $ E(\sigma_x^B|\sigma_x^A)$ and  $E(\sigma_y^B|\sigma_y^A) $ are defined in Eq. (\ref{con}). Hereinafter, we provide a comparative study between the average of conditional entropy squeezing and one-way quantum steering for some different quantum systems.

%%%%%%%%%%%%%%%%%%%%%%%%%%%
\section{Some Results and Discussion}\label{S2}
Here, we study the relationship between one-way steering and the average conditional entropy squeezing for a class of two-qubit state, which reads
\begin{equation} \label{gs}
	\hat{\rho}_{AB}=\nu |\phi\rangle \langle \phi |+ (1-\nu) |\psi\rangle \langle \psi |,
\end{equation}
where $ |\phi\rangle=\frac{|01\rangle+|10\rangle}{\sqrt{2}} $, $ |\psi\rangle=\frac{|00\rangle+|11\rangle}{\sqrt{2}} $, and $ \nu $ is the setting state parameter. The state (\ref{gs}) is maximally entangled for  $\nu=1$ and $\nu=0$, while partially entangled for $ \nu\in(0,0.5) \cup (0.5,1)$.

\begin{figure}[h!]
	\begin{center}
		\includegraphics[width=0.45\textwidth, height=4.5cm]{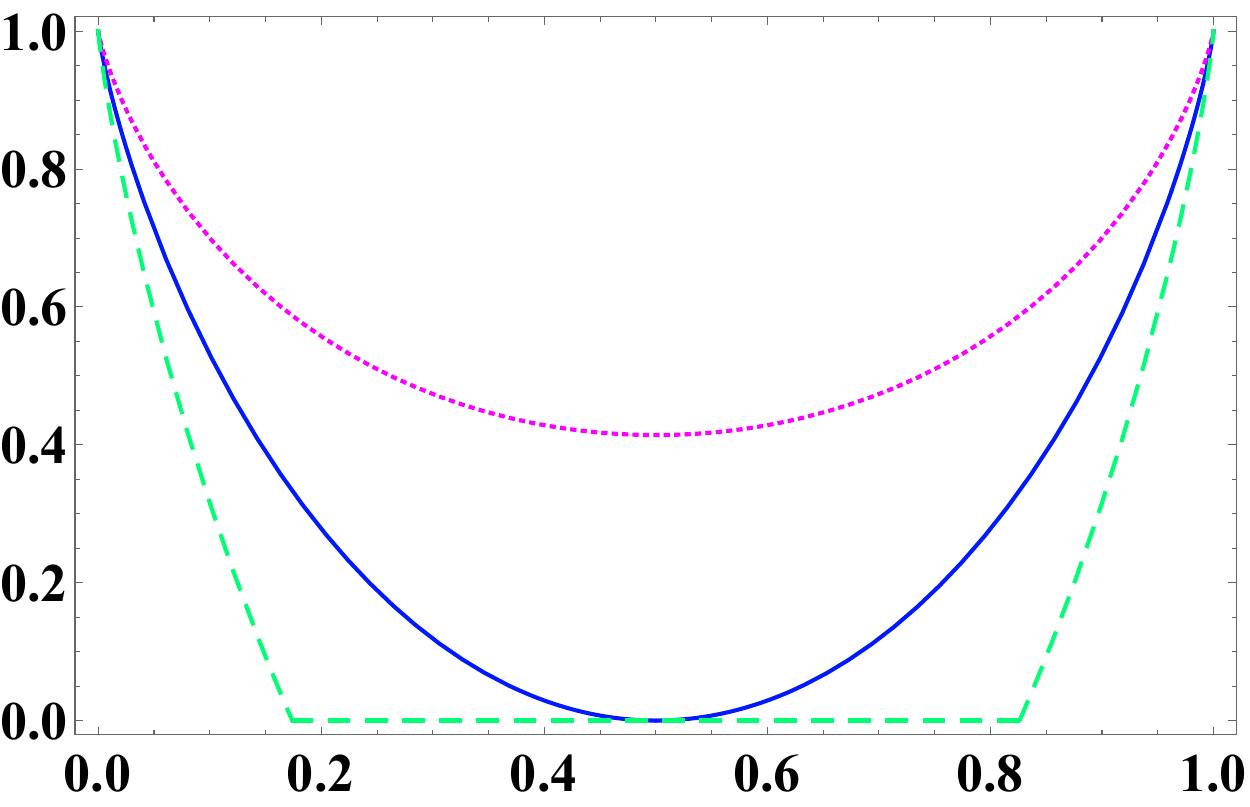}
		\put(-200,130){($ a $)}\put(-120,-11){$\nu$}\\	
		\includegraphics[width=0.45\textwidth, height=4.5cm]{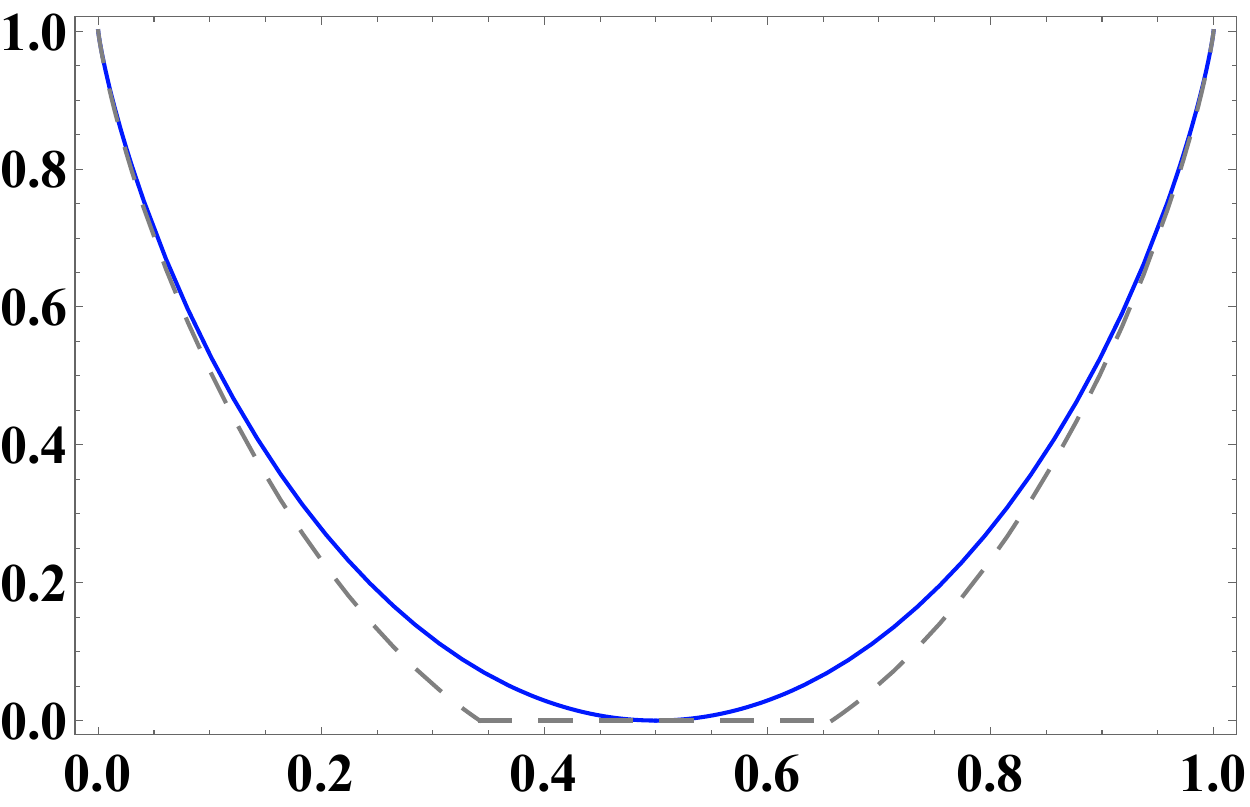}
		\put(-200,130){($ b $)}\put(-120,-11){$\nu$}
	\end{center}
	\caption{Comparative study between (a) $ E(\sigma_x^B|\sigma_x^A) $ (dotted curve),  $ E(\sigma_y^B|\sigma_y^A) $ (dashed curve), and one-way NEUR-steering $ S^{A\longrightarrow B} $ (solid curve) for the state (\ref{gs}), and (b) the average of conditional entropy squeezing $ \mathcal{Z}^{A\longrightarrow B} $ (dashed curve) and one-way NEUR-steering $ S^{A\longrightarrow B}  $ (solid curve). }
	\label{f1}
\end{figure}

In Figure \ref{f1}, we have performed a comparative analysis of the NEUR-steering  and entropy squeezing for a composite system consisting of a two-qubit state represented by Eq. (\ref{gs}). Through our analysis, we have observed interesting relationships between these two measures. Figure \ref{f1}(a) clearly demonstrates that the extent of NEUR-steering is bounded by the two quadratures of entropy squeezing. Specifically, when the NEUR-steering is maximized, we observe that the two quadratures of entropy squeezing become identical. Conversely, when the NEUR-steering is minimized, the two quadratures of entropy squeezing are separated. This finding indicates that the values of $ E(\sigma_x^B|\sigma_x^A) $ and $ E(\sigma_y^B|\sigma_y^A) $ can be considered as upper and lower bounds of NEUR-steering, respectively. Furthermore, we have investigated the average of the two quadratures of entropy squeezing  and its relation with the NEUR-steering. Figure \ref{f1}(b) illustrates that at maximally entanglement $ \nu=0 $ and $ \nu=1 $, the average of entropy squeezing ($ \mathcal{Z}^{A\longrightarrow B} $) aligns closely with the NEUR-steering $ S^{A\longrightarrow B}  $. However, at a lower degree of steering, corresponding to a partially entangled state, we observe deviations between the behaviors of NEUR-steering and the average of entropy squeezing. Nevertheless, even in these cases, $ \mathcal{Z}^{A\longrightarrow B} $ remains a reliable indicator for expressing the presence of steerability in the system.

%%%%%%%%%%%%%%%%%%%%%%%%
\subsection{Some Quantum Processes}
In this subsection, we will compare in detail the effect of some quantum processes on the functions $ \mathcal{Z}^{A\longrightarrow B} $ and  $ S^{A\longrightarrow B}  $, namely acceleration process, decoherence via a stochastic dephasing channel, and swapping process.

\subsubsection{Acceleration Process}
Let two qubits be simultaneously or separately accelerated in Rindler space. The computational basis $ \{0,1\} $ in this space for regions $ I $ and $ II $ can be defined as \cite{PhysRevA.74.032326}
\begin{equation}
	\begin{split}
	&|0_k\rangle= \cos r_k |0_k\rangle_I |0_k\rangle_{II}+ \sin r_k |1_k\rangle_I |1_k\rangle_{II},\\&
	|1_k\rangle= |1_k\rangle_I |0_k\rangle_{II},
	\end{split}
\end{equation}
where $ r_k\in [0,\pi/4] $ is the acceleration parameter of the qubit $ k=A,B $. By substituting in the state (\ref{gs}) and tracing over the degrees of the region $ II $, one can get the accelerated state as
\begin{equation}\label{acc}
	\begin{split}
		\hat{\rho}_{AB}^{acc}&= \mathcal{A}_{11} |00\rangle \langle 00| + \mathcal{A}_{22}|01\rangle \langle 01|+ \mathcal{A}_{33}|10\rangle \langle 10|\\&+ \mathcal{A}_{44} |11\rangle \langle 11| +( \mathcal{A}_{14} |00\rangle \langle 11|+\mathcal{A}_{23} |10\rangle \langle 01|+h.c.),
	\end{split}
\end{equation}
 where
\begin{equation}
	\begin{split}
		&\mathcal{A}_{11}= \frac{\nu}{2} \cos^2 r_a \cos^2 r_b ,\ \mathcal{A}_{22}= \cos^2 r_a (\frac{\nu}{2} \sin^2r_b +\frac{1-\nu}{2}),\\
		&\mathcal{A}_{33}=\cos^2 r_b (\frac{\nu}{2} \sin^2r_a + \frac{1-\nu}{2}), \\&
		 \mathcal{A}_{44}=  \sin^2 r_a (\frac{\nu}{2} \sin^2r_b + \frac{1-\nu}{2})+  \frac{1-\nu}{2} \sin^2r_b +\frac{\nu}{2}, \\
		  &\mathcal{A}_{14}=\frac{\nu}{2}\cos r_a \cos r_b, \  \mathcal{A}_{23}=\frac{1-\nu}{2}\cos r_a \cos r_b.
	\end{split}
\end{equation}

\begin{figure}[h!]
	\begin{center}
		\includegraphics[width=0.45\textwidth, height=4.5cm]{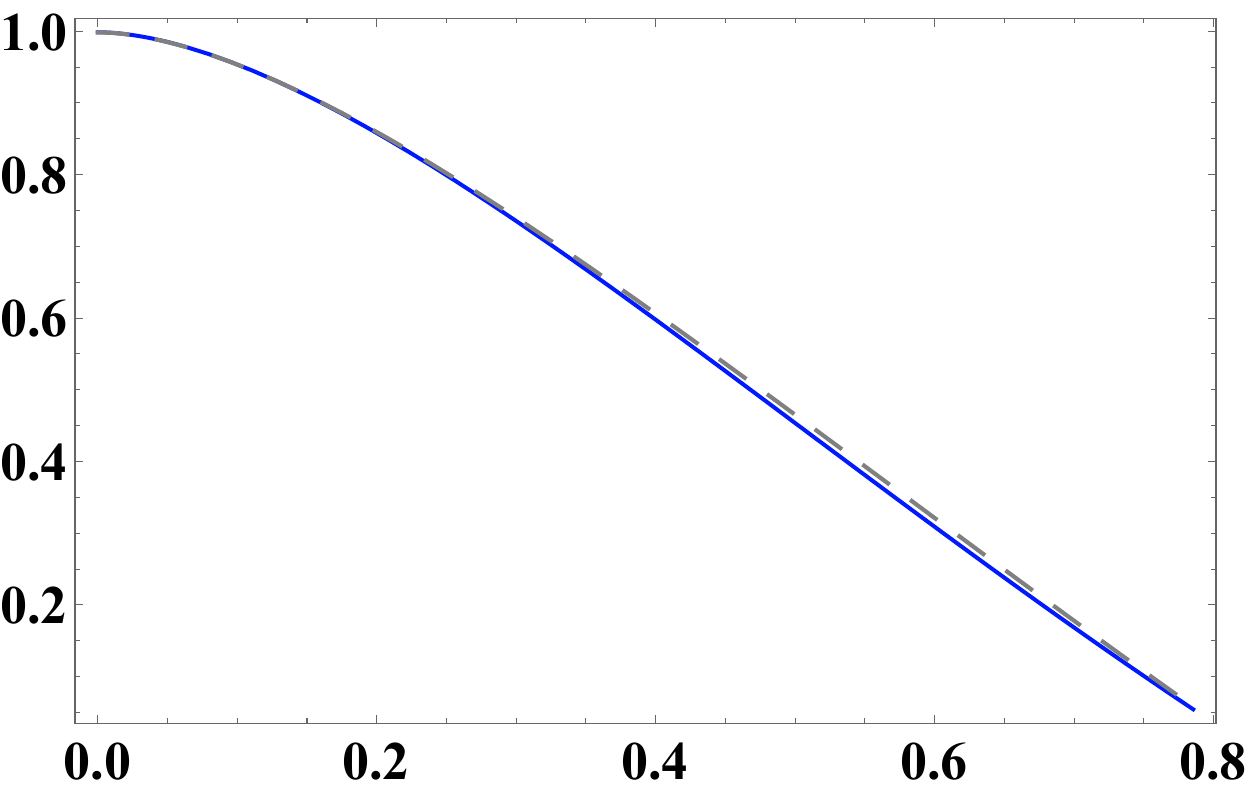}
		\put(-200,130){($ a $)}\put(-120,-11){$r$}\\	
		\includegraphics[width=0.45\textwidth, height=4.5cm]{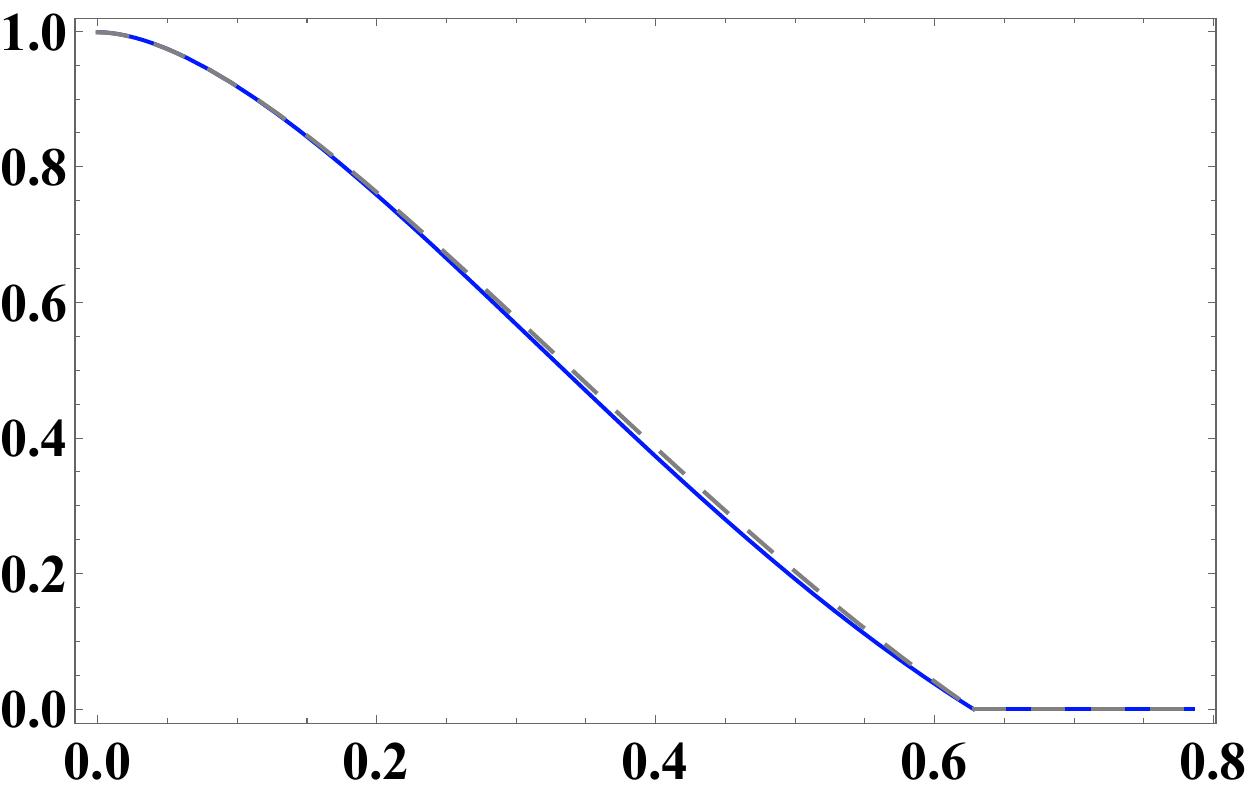}
		\put(-200,130){($ b $)}\put(-120,-10){$r$}
	\end{center}
	\caption{Average of conditional entropy squeezing $ \mathcal{Z}^{A\longrightarrow B} $ (dashed curve) and one-way NEUR-steering $ S^{A\longrightarrow B}  $ (solid curve) for accelerated state (\ref{acc}) with $\nu=1$. (a) $ r_a=r $, $ r_b=0 $ and (b) $ r_a= r_b=r $.}
	\label{f2}
\end{figure}

In Figure \ref{f2}, we present an investigation into the impact of the accelerated process on a two-qubit state. Specifically, we aim to explore the relationship between the NEUR-steering and average entropy-squeezing measures under this process. We assumed that the two-qubit state is maximally entangled with $\nu=1$ as a fixed parameter. In Figure \ref{f2}(a), we observe that when only one qubit is accelerated with $ r_a=r $ and $ r_b=0 $, the NEUR-steering is maximized at lower values of the acceleration parameter. As the acceleration parameter increases, we see a decrease in the degree of steering. Interestingly, we note that the NEUR-steering and the entropy squeezing are identical across different values of the acceleration parameter $ r $. This indicates a consistent relationship between these measures regardless of the acceleration applied to the system. On the other hand, when both qubits are accelerated simultaneously with $ r_a= r_b=r $, Figure \ref{f2}(b) reveals an intriguing trend. As the acceleration parameter increases, the rate of decrease in steering becomes more pronounced. This finding suggests that accelerating two qubits simultaneously increases the suppression of steering. However, it is important to note that despite this trend, the two measures, namely NEUR-steering and entropy squeezing, exhibit little variations with respect to the acceleration parameter.

%%%%%%%%%%%%%%%%%%%%%%%%%%%
\subsubsection{Noisy Channel Process}
To examine the two functions (\ref{sab}) and (\ref{ZZ}) under noisy channel models, we can express the temporal density operator in terms of Kraus operators as
\begin{equation}
	\hat{\rho}_{AB}(t)= \sum_{i,j} K_i^A(t)  K_j^B(t) \hat{\rho}_{AB}(0) (K_i^A (t)  K_j^B (t))^\dagger,
\end{equation}
here $ \hat{\rho}_{AB}(0) $ is defined in Eq. (\ref{gs}), while $ K_i^k(t)$ and $ K_j^k(t)$ with $k=A, B $ are the time-dependent Kraus operators for different noise channels. For example, we use the Kraus operators of amplitude-damping noise, which are defined by \cite{thapliyal2017quantum}

\begin{equation}
	K_1 (t)=|0\rangle\langle 0|+\sqrt{1-P(t)} |1\rangle\langle 1|, \quad  K_2 (t)=\sqrt{P(t)} |0\rangle\langle 1|,
\end{equation}
where $ P(t)= e^{-g t}\left[\cos(\frac{\lambda t}{2})+ \frac{g}{\lambda} \sin(\frac{\lambda t}{2}) \right]^2 $ with  $\lambda=\sqrt{ g(2\gamma - g)}$. Herein, $ g $ is a decay rate which depends on the reservoir correlation time, and $ \gamma $  is the coupling strength related to qubit relaxation time.

Likewise, the Kraus operators for purely dephasing noise channels can be defined as \cite{yu2010entanglement}
\begin{equation}
	K_1 (t)=|0\rangle\langle 0|+P(t) |1\rangle\langle 1|, \quad K_2(t)=\sqrt{1-P^2(t)} |1\rangle\langle 1|,
\end{equation}
where $$ P(t)=\exp\left\{-\dfrac{\gamma}{2} \left(t+g^{-1} [\exp(-g t)-1]\right)\right\}. $$

\begin{figure}[h!]
	\begin{center}
		\includegraphics[width=0.45\textwidth, height=4.5cm]{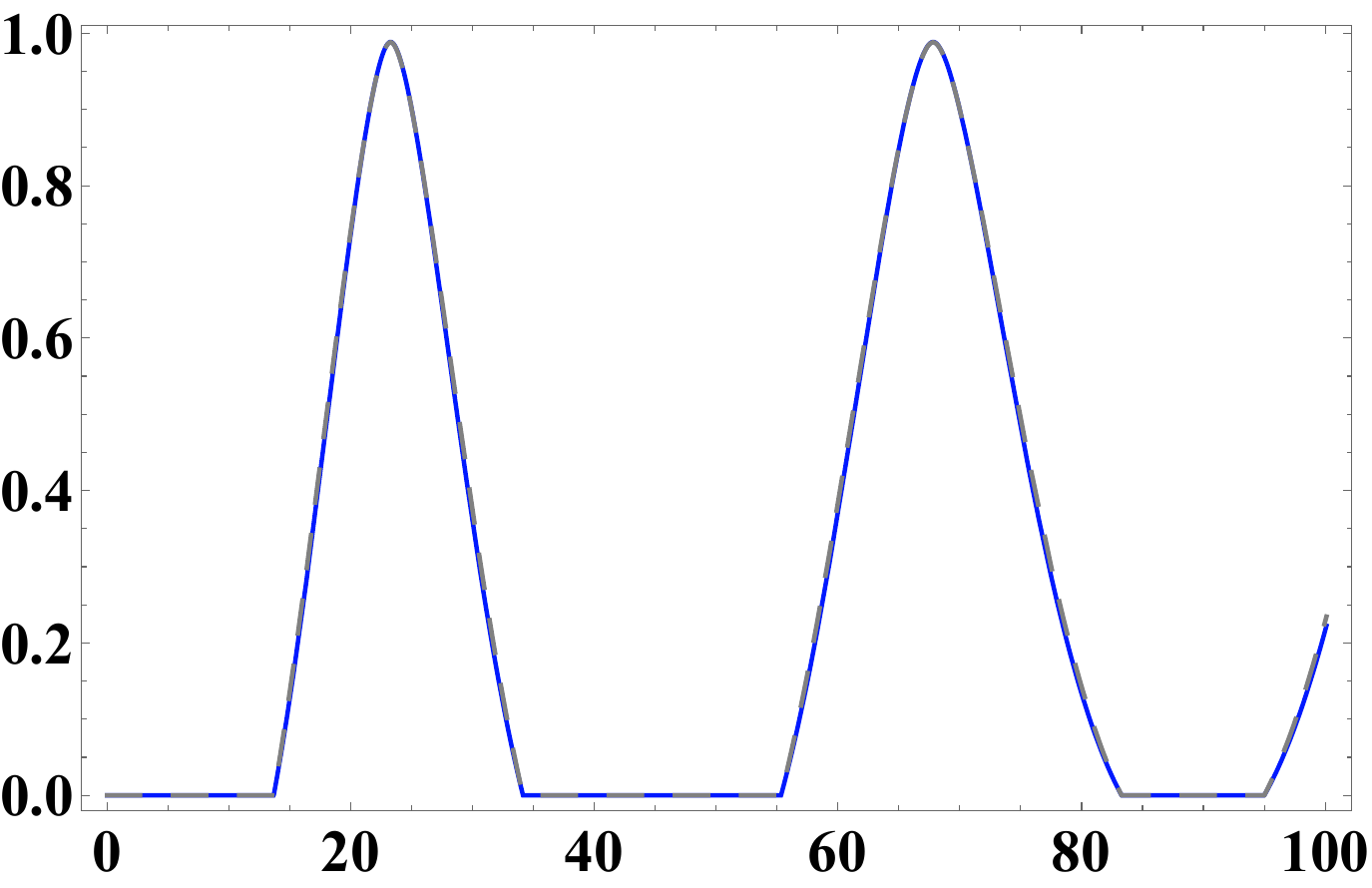}
		\put(-200,130){($ a $)}\put(-120,-11){$\gamma t$}\\	
		\includegraphics[width=0.45\textwidth, height=4.5cm]{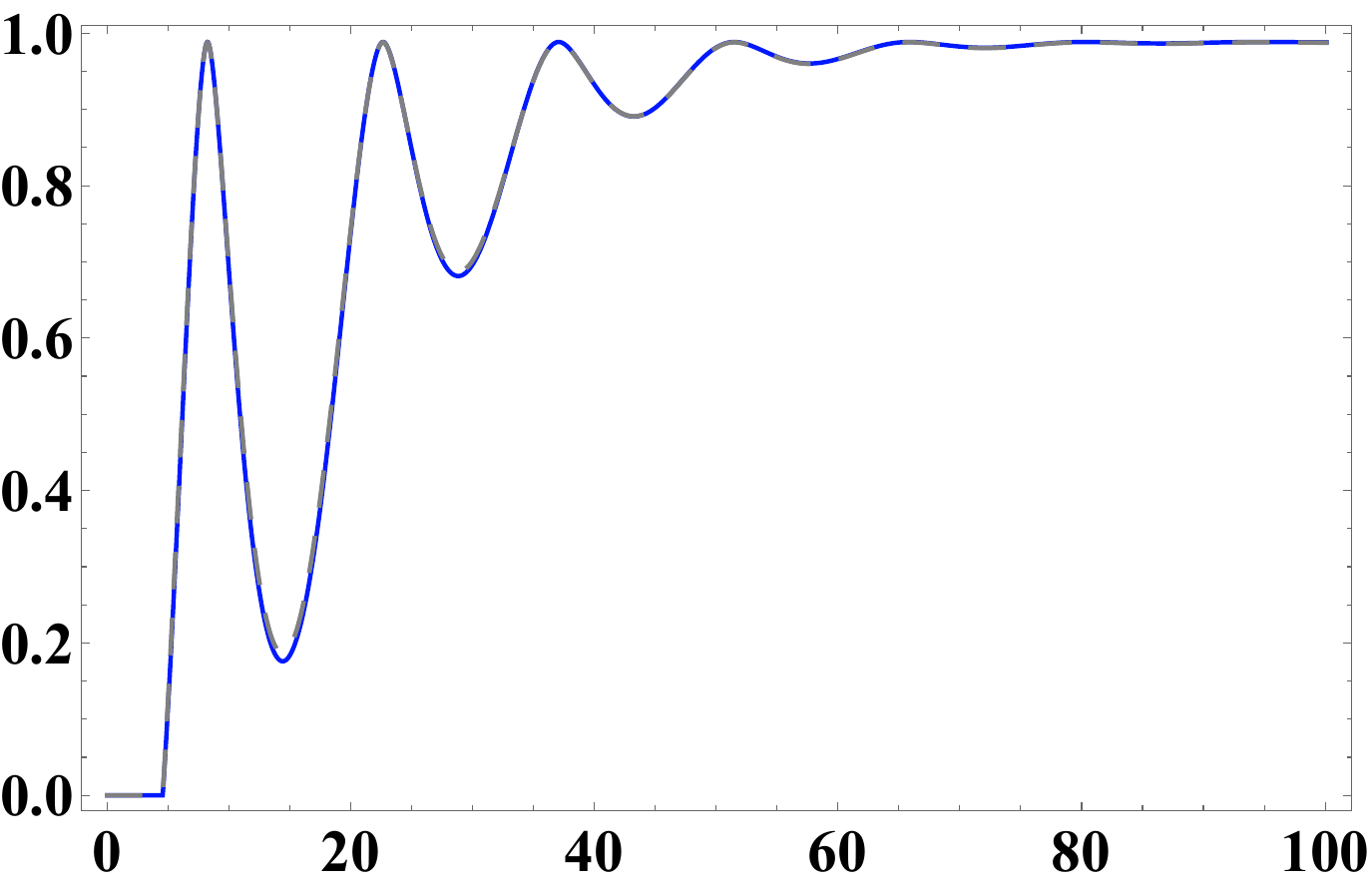}
		\put(-200,130){($ b $)}\put(-120,-10){$\gamma t$}\\
		\includegraphics[width=0.45\textwidth, height=4.5cm]{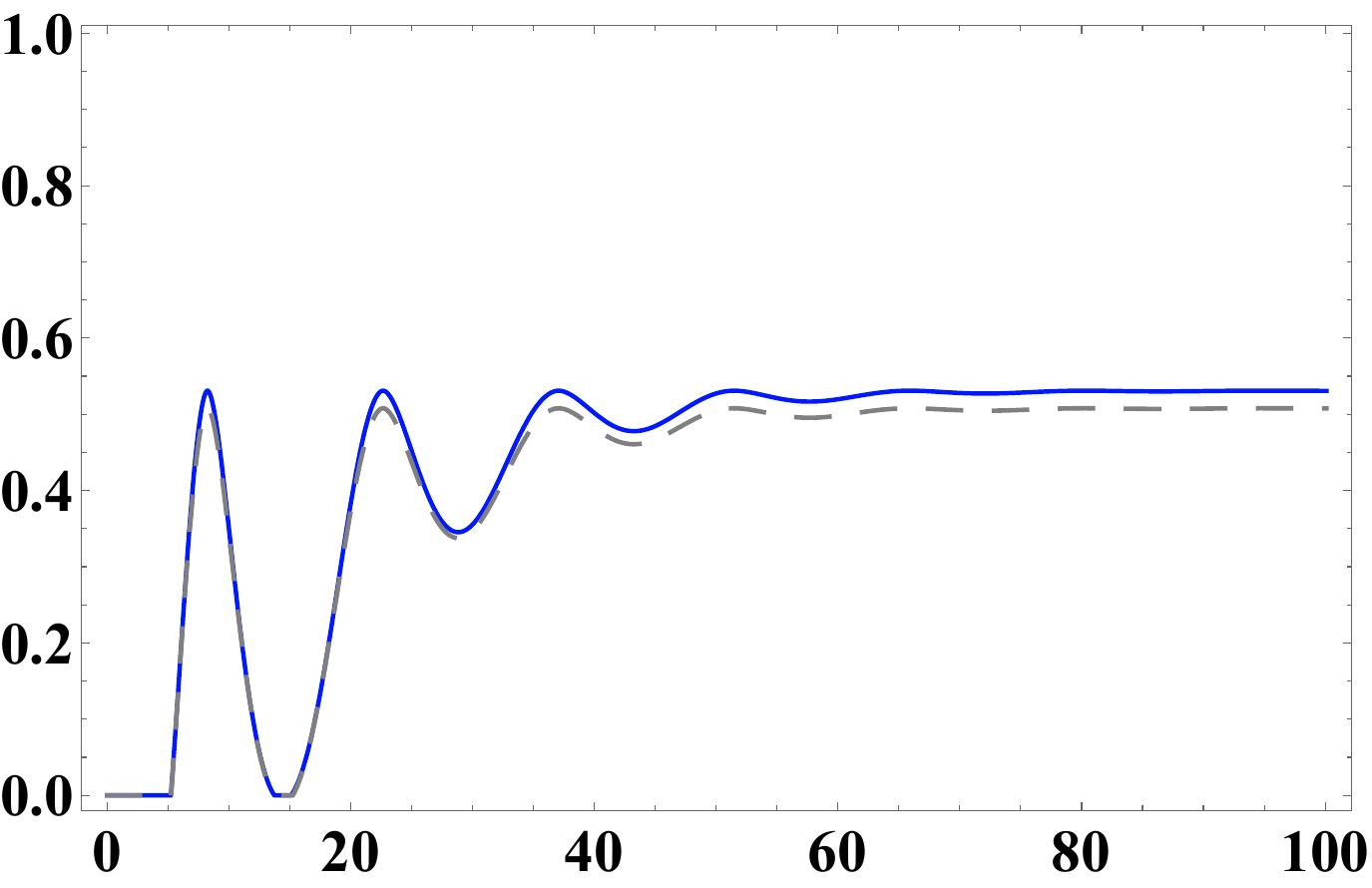}
		\put(-200,130){($ c $)}\put(-120,-10){$\gamma t$}
	\end{center}
	\caption{Average of conditional entropy squeezing $ \mathcal{Z}^{A\longrightarrow B} $ (dashed curve) and one-way NEUR-steering $ S^{A\longrightarrow B}  $ (solid curve) for amplitude-damping noise, where (a) $\nu=1$, $ g=0.01 $,  (b) $\nu=1$, $g=0.1 $ and (c) $\nu=0.1$, $g=0.1 $.}
	\label{f3}
\end{figure}
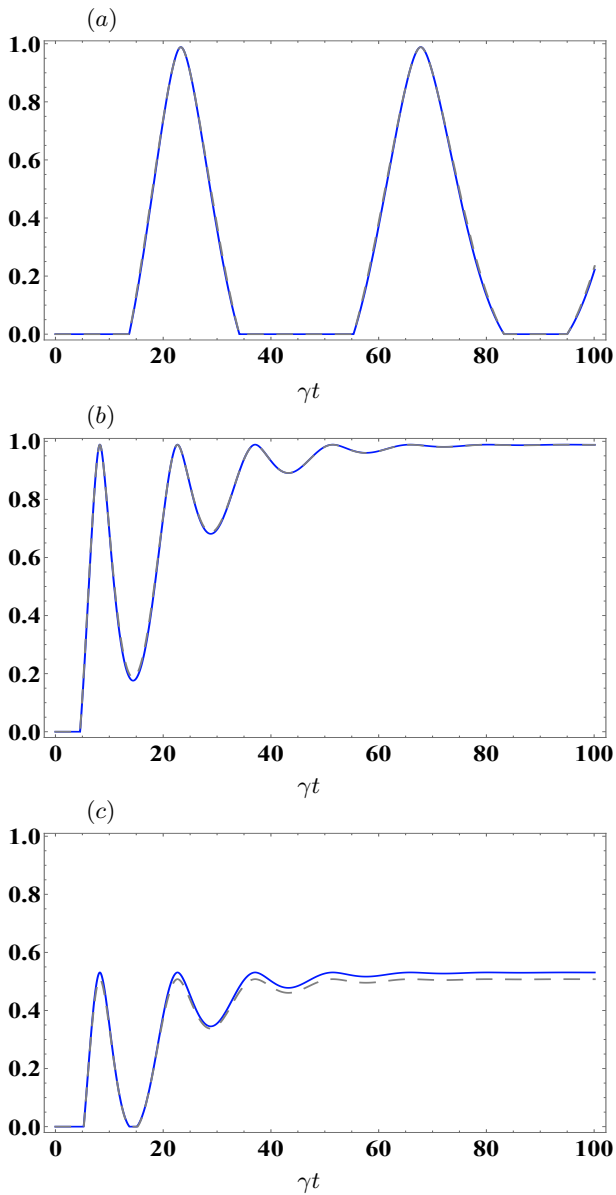
Figure \ref{f3} presents a comprehensive examination of the impact of amplitude-damping noise on the degree of steerability, with the NEUR-steering and entropy squeezing serving as the quantifying measurements. In Figure \ref{f3}(a), we observe that, by selecting a small value for the damping rate ($ g=0.01 $), the NEUR-steering oscillates between its maximum and lower bounds. This oscillatory behaviour is consistent with the properties of steerability under the influence of amplitude-damping noise. It is noteworthy that the two measures, NEUR-steering and entropy squeezing, coincide perfectly with the scaled time parameter. This convergence of the two measurements further reinforces their equivalence in capturing the system's dynamics. Moving to Figure \ref{f3}(b), we examine the effect of increasing decay rates within the range of $g$ (specifically with $ g=0.1 $ and maximally entangled state with $ \nu=1 $). As the scaled time escalates $ \gamma t $, we gradually observe the NEUR-steering oscillating with an increase in the upper bounds. On the other hand,  in the case of a partially entangled state with a parameter value of $ \nu=0.1 $, it can be observed from Figure \ref{f3}(c) that a clear separation between the two measures over time. Furthermore, as the scaled time increases, the upper bounds of the steering degree exhibit a decrease in value. Remarkably, the maximum bounds of steering continue to expand as time progresses, indicating a growing influence of the amplitude-damping noise on the steerability of the system. Just like in the previous case, the NEUR-steering and average entropy-squeezing measurements in this scenario remain parallel, underscoring their identical nature. This consistent agreement can be attributed to the initial state being maximally entangled.

\begin{figure}[h!]
	\begin{center}
		\includegraphics[width=0.45\textwidth, height=4.5cm]{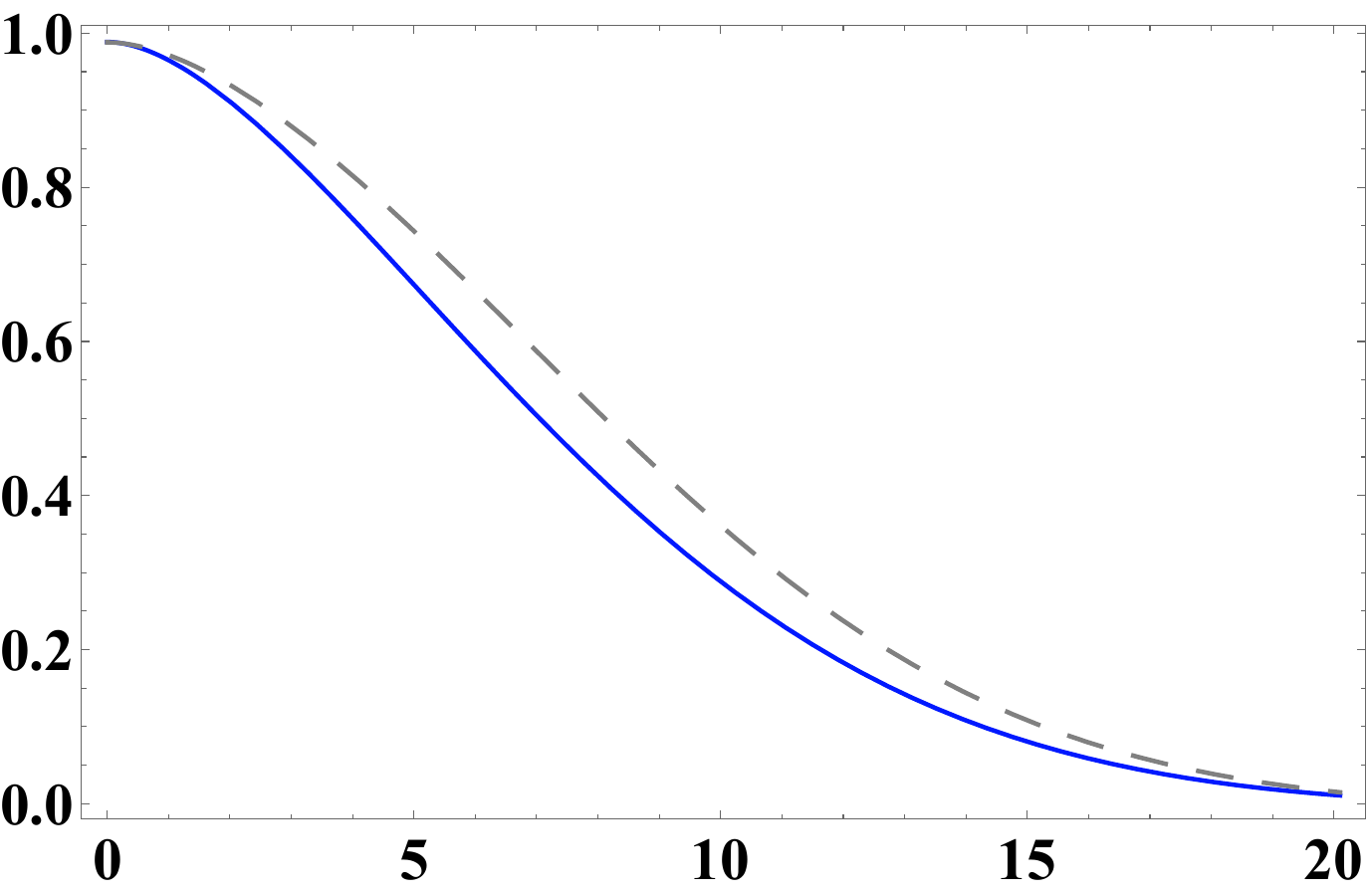}
		\put(-200,130){($ a $)}\put(-120,-11){$\gamma t$}\\	
		\includegraphics[width=0.45\textwidth, height=4.5cm]{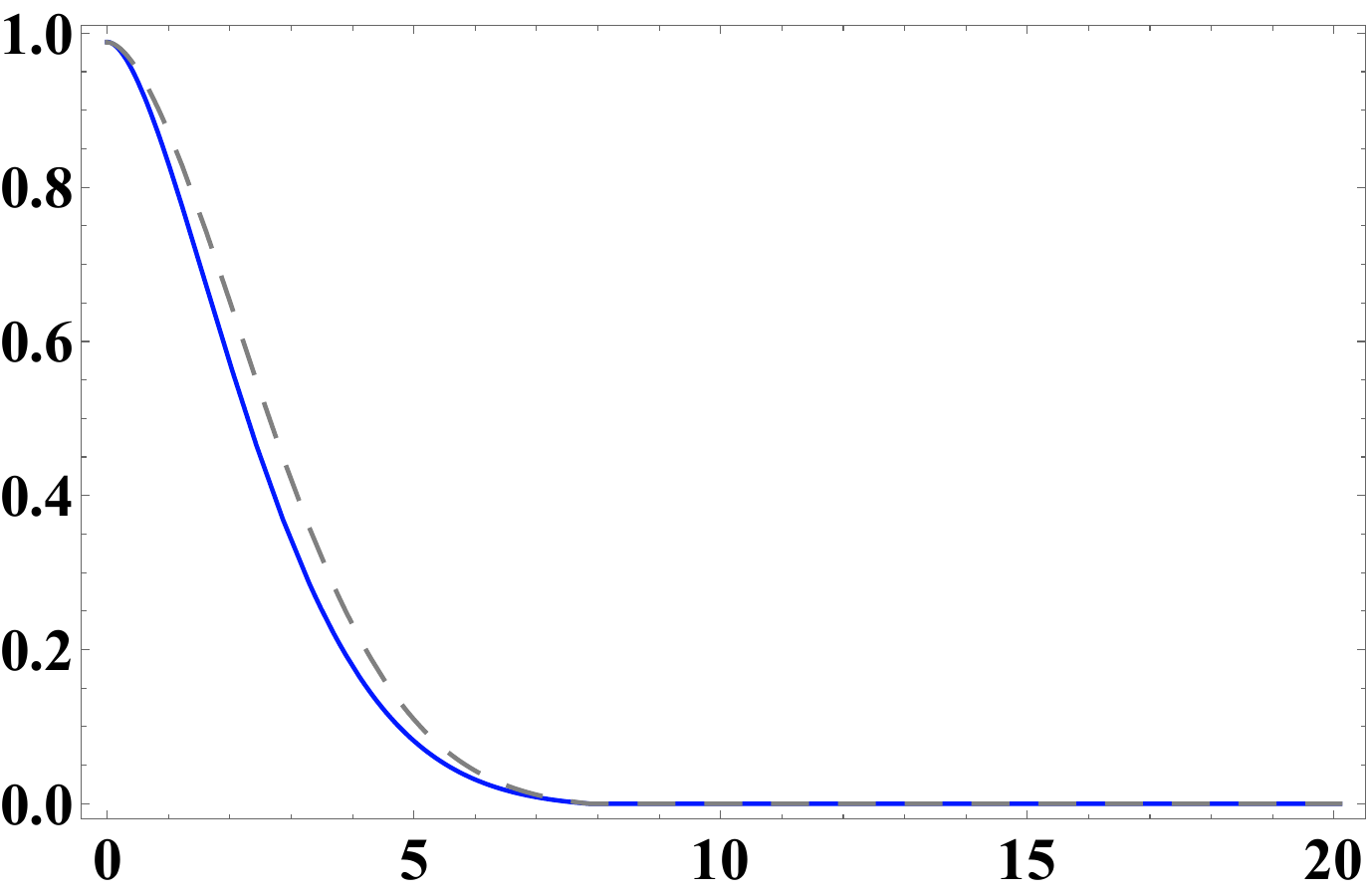}
		\put(-200,130){($ b $)}\put(-120,-10){$\gamma t$}\\
		\includegraphics[width=0.45\textwidth, height=4.5cm]{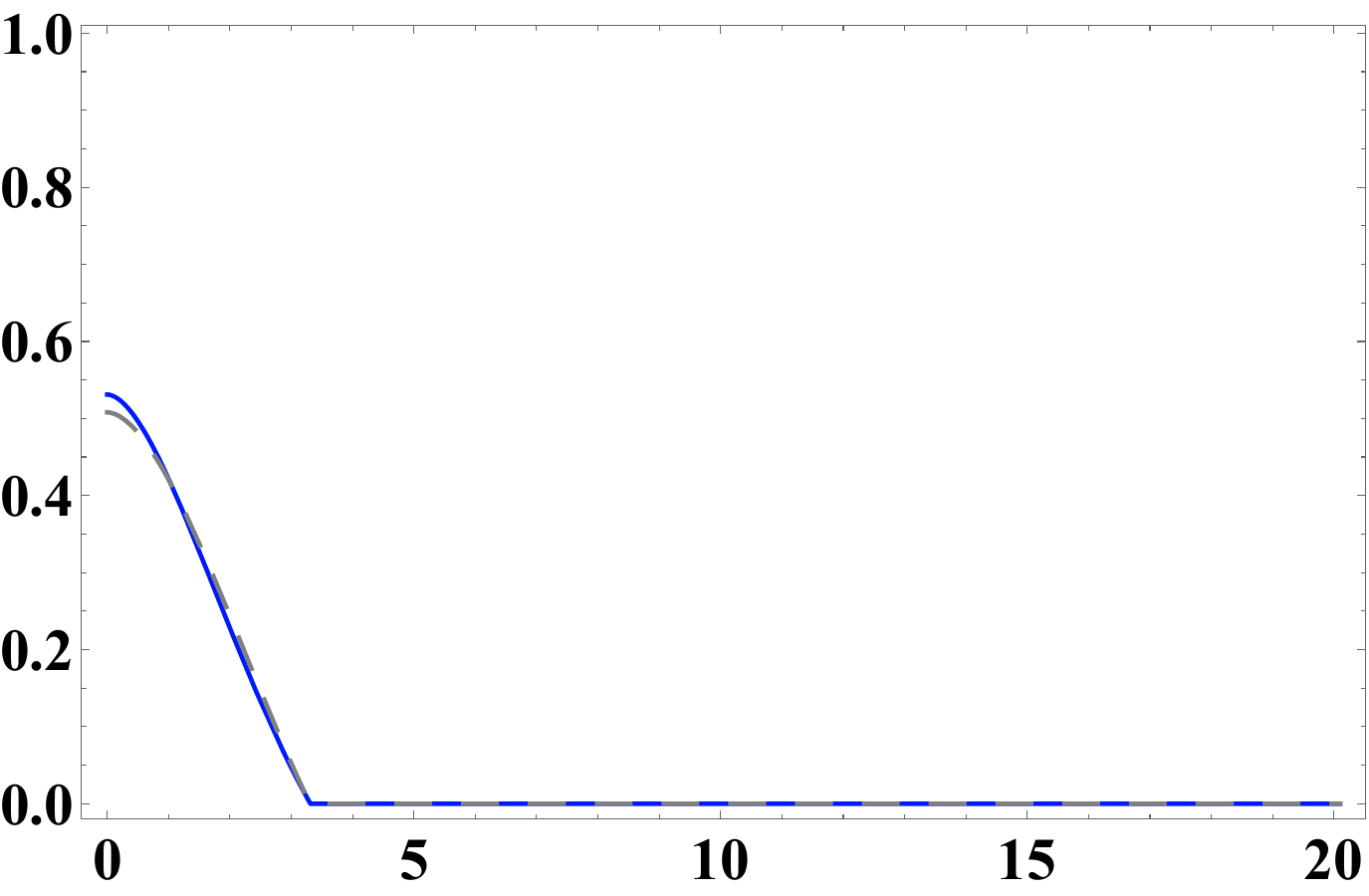}
		\put(-200,130){($ c $)}\put(-120,-10){$\gamma t$}
	\end{center}
	\caption{Average of conditional entropy squeezing $ \mathcal{Z}^{A\longrightarrow B} $ (dashed curve) and one-way NEUR-steering $ S^{A\longrightarrow B}  $ (solid curve) under purely dephasing channel, where (a) $\nu=1$, $ g=0.01 $,  (b) $\nu=1$, $g=0.1 $ and (c) $\nu=0.1$, $g=0.1 $.}
	\label{f4}
\end{figure}
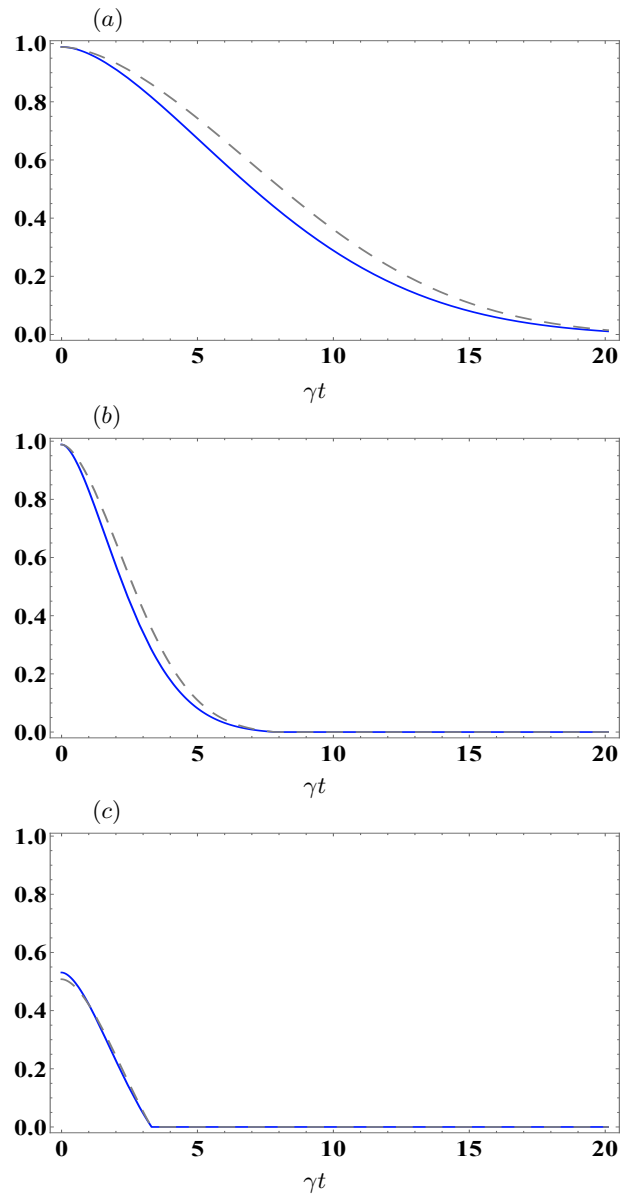

Figure \ref{f4} provides a detailed analysis of the effect of dephasing noise on the degree of steering, utilizing the NEUR-steering and entropy squeezing as the measurement criteria. Moreover, the comparison between these two measures under the influence of the dephasing noise channel is examined. In the first, we consider an initial state that is maximally entangled with $ \nu=1 $. Initially, we note that while the entropy squeezing and NEUR-steering measures display similarities in their general behavior, they are not entirely identical as the scaled time progresses. At the onset, both measures exhibit their maximum bounds. However, as time increases, we observe a notable distinction between the maximum bounds of entropy squeezing and NEUR-steering. Interestingly, the maximum bounds of entropy squeezing surpass those of NEUR-steering. This disparity suggests that the influence of dephasing noise imposes a more pronounced impact on the entropy squeezing measure compared to NEUR-steering. In the context of partial entanglement $ \nu=0.1 $ with $ g=0.1 $, our observations indicate a rapid decay of steering and a decrease in the upper bounds of steering during the initial stages of the interaction. This suggests that partial entanglement has a significant impact on the dynamics of steering. Furthermore, it is important to note that the time evolution of the steered system is influenced by the degree of entanglement. As the entanglement decreases, the decay of steering becomes more pronounced, implying a decreasing ability to remotely control and influence the entangled particles.

Notably, despite the difference in the maximum bounds between the two measures, they still portray parallel trends. As the scaled time continues to grow, we witness a decrease in the degrees of steering for both measures. This observation implies that the detrimental effects of dephasing noise manifest as a reduction in the correlation between the entangled subsystems. Besides, it is evident that increasing the damping rate of the dephasing channel exacerbates this decreasing effect on the steering degrees.

%%%%%%%%%%%%%%%%%%%%%%%%%%%%%%%
\subsubsection{Swapping Process}
Let us consider two different sources, $ S_{12} $ and $ S_{34} $, which generate pairs of two-qubit state $ \rho_{12} $ and $ \rho_{34} $, respectively. Qubits 1 and 4 are far apart, while qubits 3 and 2 remain close. The swapping process is aimed to measure the amount of quantum NEUR-steering between qubits 1 and 4 by performing a joint Bell measurement on qubits 2 and 3. The post-measurement state $ \rho_{14} $ is calculated by \cite{rosario2022quantum}
\begin{equation}
	\rho_{14}=Tr_{23} \bigg[\frac{M_i. \rho_{1234}. M_i^\dagger}{Tr[M_i. \rho_{1234}. M_i^\dagger]}\bigg],
\end{equation}
where $ \rho_{1234}=\rho_{AB} \otimes \rho_{AB}$, such that the two sources generate the state $ \rho_{AB} $ defined in Eq. (\ref{gs}). Moreover, $ M_i = I_2 \otimes|\Phi_i\rangle \langle \Phi_i|\otimes I_2$ and $ |\Phi_i\rangle $  stand for the usual four Bell states.

\begin{figure}[h!]
	\begin{center}
		\includegraphics[width=0.45\textwidth, height=4.5cm]{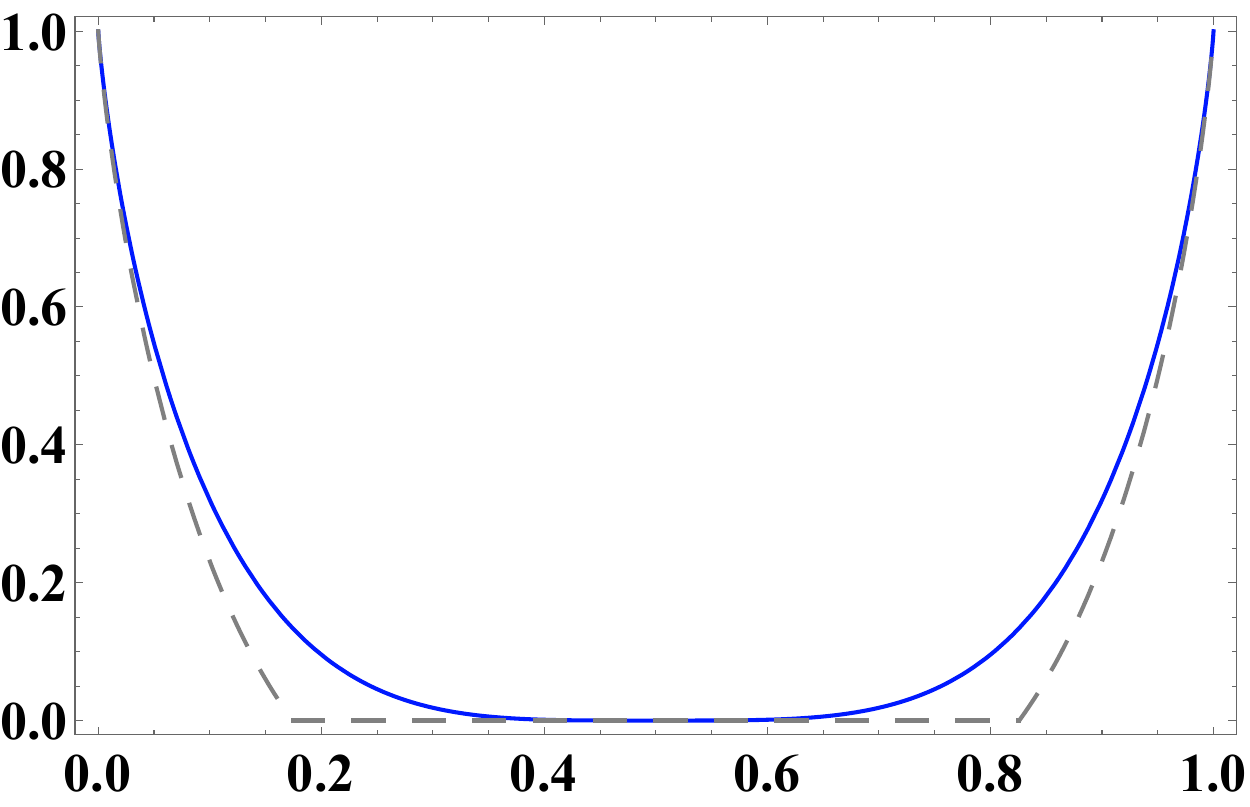}
		%\put(-125,130){ }
  \put(-120,-11){$\nu$}
	\end{center}
	\caption{Average of conditional entropy squeezing $ \mathcal{Z}^{A\longrightarrow B} $ (dashed curve) and one-way NEUR-steering $ S^{A\longrightarrow B}  $ (solid curve)  under swapping process with $ |\Phi_i\rangle=|\psi\rangle $. }
	\label{f5}
\end{figure}
Finally, Figure \ref{f5} focuses on analyzing the effects of the swapping process on the behavior of  NEUR-steering and conditional entropy squeezing concerning the state parameter $ \nu $. The post-measurement state $ \rho_{14} $ at $ |\Phi_i\rangle=|\psi\rangle $ is studied to understand how the swapping process influences the behavior of the quantum system. The findings indicate that the steerability degree significantly decreases when the two-qubit state is initially in a partially entangled state. Furthermore, when compared to Figure \ref{f1}, it is noticeable that the unsteerable region expands during the swapping process. However, when the two-qubit state is maximally entangled, the two measures are equal. Hence, conditional entropy squeezing serves as an excellent indicator of the level of NEUR-steering present under this process.

%%%%%%%%%%%%%%%%%%%%%%%%%
\section{Conclusion}\label{S3}
We have proposed a new method for quantifying one-way quantum NEUR-steering in an arbitrary two-qubit system using the average of conditional entropy squeezing. We derived the explicit analytical expressions of NEUR-steering and conditional entropy squeezing. A comparative analysis of the two measures was conducted on a free maximally mixed two-qubit state, either restricted two-qubit state by using acceleration, noisy channels, or swapping processes.

For the free maximally mixed two-qubit state, our results highlight the interconnectedness between NEUR-steering and entropy squeezing. We demonstrated that the quadratures of entropy squeezing serve as bounds for NEUR-steering, which represented the upper and lower limits. Additionally, we established the average of entropy squeezing as a valuable indicator of steerability, particularly at a maximally entangled state.

The effects of the accelerated process on a two-qubit state have been demonstrated. We observed that the behavior of NEUR-steering and entropy squeezing depends on whether one or both qubits are accelerated. In the former case, the degree of NEUR-steering decreases with increasing acceleration, while in the latter case, the rate of decrease in steering is amplified. Nonetheless, despite these variations, the measures of NEUR-steering and entropy squeezing remain stable and invariant with respect to the acceleration parameter.

Under the amplitude-damping noise, our results showed that under a specific small damping rate, the NEUR-steering experiences oscillatory behavior. By allowing the decay rate to increase, the NEUR-steering experiences oscillatory behavior with expanding boundaries. Notably, the NEUR-steering and entropy-squeezing measurements remained indistinguishable throughout these processes, further validating their correlation. On the other hand, the effect of dephasing noise on the degree of steering has been evaluated. While the two measures differ in terms of their maximum bounds, they exhibit similar overall trends. Specifically, as the scaled time increases, both measures demonstrate a decrease in the degree of steering. This decreasing effect is amplified by enhancing the damping rate of the dephasing channel.

Finally, the swapping process significantly diminishes the steerability degree when the initial state of the two-qubit is partially entangled.  For maximally entangled states, the two measures coincide.

In conclusion, it is evident that entropy squeezing serves as a measure of steering primarily in the context of a maximally entangled system. However, when considering partially entangled states or situations where entanglement is constrained by external factors, entropy squeezing remains a highly reliable indicator of steering. By assessing the degree of entropy squeezing in such scenarios, valuable insights can be gained regarding the presence and extent of quantum steering. Entropy squeezing is a tool for indicating and understanding quantum steering, even in cases where maximal entanglement may not be achieved.
\\
\\
\\
\textbf{Data availability}\\
All data generated during this study are included in this paper.\\
\\
\\
\textbf{Competing interests}\\
The authors declare no competing interests.\\
\\
\\
\textbf{ORCID iDs}\\
\noindent A-S. F. Obada\\ \href{https://orcid.org/0000-0001-5862-7365}{https://orcid.org/0000-0001-5862-7365}\\
M. Y. Abd-Rabbou \\\href{https://orcid.org/0000-0003-3197-4724}{https://orcid.org/0000-0003-3197-4724}\\
Saeed Haddadi \\\href{https://orcid.org/0000-0002-1596-0763}{https://orcid.org/0000-0002-1596-0763}
\\
\\
\\
\\
\\
\\

\vfill
\bibliographystyle{unsrt}
\bibliography{bm}

%%%%%%%%%%%%%%%%%%%%%%%%%%%%%%%%%%%%%%%%%%%%%%%%%%%%%%%%%%%%%%%%%%%%%
%%%%%%%%%%%%%%%%%%%%%%%%%%%%%%%%%%%%%%%%%%%%%%%%%%%%%%%%%%%%%%%%%%%%

\end{document}